\documentclass[useAMS,usenatbib]{mn2e}
\usepackage{epsfig}

\title[Gas in the $z\sim8$ GRB Host Galaxy]{Constraining the Molecular Gas in the Environs of a $\mathbf{z\sim8}$ Gamma Ray Burst Host Galaxy}
\author[E R Stanway et al.]{Elizabeth R.~Stanway$^{1}$\thanks{email: E.R.Stanway@Bristol.ac.uk}, Malcolm N. Bremer$^{1}$, Nial R.~Tanvir$^{2}$, Andrew J.~Levan$^{3}$, \newauthor Luke J.~M.~Davies$^{1}$
\\$^1$ H H Wills Physics Laboratory, Tyndall Avenue, Bristol, BS8 1TL, UK
\\$^2$ Department of Physics and Astronomy, University of Leicester, University Road, Leicester, LE1 7RH, UK
\\$^3$ Department of Physics, University of Warwick, Coventry, CV4 7AL, UK}
\begin{document}

\date{Accepted 2010 August 15. Received 2010 August 06; in original form 2010 June 07}

\pagerange{\pageref{firstpage}--\pageref{lastpage}} \pubyear{2010}

\maketitle

\label{firstpage}

\begin{abstract}
GRB\,090423 is the most distant spectroscopically-confirmed source
observed in the Universe. Using observations at 37.5\,GHz, we place
constraints on molecular gas emission in the CO(3-2) line from its
host galaxy and immediate environs. The source was not detected
either in line emission or in the rest-frame 850\,$\mu$m continuum,
yielding an upper limit of $S_\mathrm{8mm}$=9.3\,$\mu$Jy and
M(H$_2$)$<4.3\times10^{9}$\,M$_\odot$ (3\,$\sigma$), applying standard
conversions. This implies that the host galaxy of GRB\,090423 did not
possess a large reservoir of warm molecular gas but was rather modest
either in star formation rate or in mass. It suggests that this was
not an extreme starburst, and hence that gamma ray bursts at high
redshift trace relatively modest star formation rates, in keeping with
the behaviour seen at lower redshifts. We do, however, identify a
millimetre emission line source in the field of
GRB\,090423. Plausible interpretations include a CO(1-0) emitting
galaxy at $z=2.1$, CO(2-1) at $z=5.2$ and CO(3-2) at $z=8.3$. Efforts
to identify a counterpart for the molecular line emitter and to
further characterise this source are continuing.
\end{abstract}


\section{Introduction}
\label{sec:intro}

Gamma Ray Bursts (GRBs) are amongst the most useful probes of the
early Universe. Their utility is two-fold, both in yielding
information about their host galaxy and immediate environment, and as
beacons, marking out mass and structure at the highest redshifts
\citep{2000ApJ...536....1L}. Given their high energies, they can
potentially be observed out to $z=15$ or further with current
instruments, but redshift determination for these high redshift bursts
is challenging, requiring deep integrations with optical and
near-infrared telescopes in the hours after a source is
detected. Despite this, the few known $z>5$ sources have provided
valuable evidence for the properties of dust
\citep{2007ApJ...661L...9S} and molecular clouds
\citep{2007ApJ...654L..17C} at these early times and for the column
density of neutral hydrogen in their
environment \citep{2007ApJ...669....1R,2006PASJ...58..485T}.

While GRBs do not require massive host galaxies
\citep{2009ApJ...691..182S,2010MNRAS.tmp..479S}, they do require the
presence of significant quantities of gravitationally collapsed and
processed baryonic matter in the form of young stars. At high
redshifts, this condition is believed to be sufficiently rare to make
them likely markers of overdense regions in the cosmic web and hence a
measure of the power of small scale density fluctuations
\citep{2005ApJ...623....1M}. Using such beacons may be the only way to
identify and characterize star forming regions during
cosmic reionization - the epoch during which the first stars heated
and ionised neutral hydrogen atoms in the intergalactic medium.

Gamma Ray Burst (GRB) 090423 at $z=8.23^{+0.06}_{-0.07}$ is the most
distant known object with a well-defined, spectroscopically-confirmed
redshift \citep{2009Natur.461.1254T,2009Natur.461.1258S}. Occurring
just 600\,Myr after the Big Bang, it likely probes the tail end of the
epoch of reionization. Its afterglow and host galaxy provide an
unprecedented opportunity to probe the properties of a star forming
galaxy at these early times, and it is likely to be the archetype for
future observations on similarly distant GRBs \citep[see the Decadal
Survey Whitepaper by][for discussion]{2009astro2010S.199M}.

Crucially, the properties of GRB\,090423 are consistent with the bulk
of the GRB population \citep{2009Natur.461.1254T}, lacking the long duration and extreme energy
output that might be expected from a pop III progenitor
\citep{2010ApJ...715..967M}. As such, this burst must mark out a
region in which star formation is already well under way, and in which
the interstellar and intergalactic medium has already undergone
processing. For this to have occurred at such high redshifts requires
 this region to have collapsed early in the history of the Universe,
implying that it is highly overdense relative to the volume average at
early times. By $z=8$, it is possible for dark matter haloes as
massive as a few $10^{12}$\,M$_\odot$ to have collapsed, albeit in small
numbers \citep{2002MNRAS.336..112M}. While GRB\,090423 is not
coincident with a known AGN, it could plausibly lie either in a
massive halo, or in the mass overdensity associated with such a halo.

Interpretation of high redshift bursts, particularly in the context of
evolutionary trends such as the cosmic star formation rate history and
inferences made from it
\citep[e.g.][]{2009ApJ...705L.104K,2009MNRAS.400L..10W,2010MNRAS.401.2561W},
requires an understanding of the burst environment and how it too
changes with redshift. At low redshift, GRBs are not unbiased tracers
of star formation, but are seen to occur preferentially in low mass,
metal poor galaxies
\citep{2009ApJ...691..182S,2010MNRAS.tmp..479S}. Semi-analytic models,
applied to large simulations such as the Millennium Simulation, suggest
that GRB host galaxy halo mass is largely independent of redshift to
$z>6$ \citep{2010arXiv1005.4036C}, but observational evidence for this
is sparse. \citet{2009ApJ...705L.104K} have suggested 
that the ratio between gamma ray burst rate and star formation density
scales as $(1+z)^{1.2}$, possibly due to evolution in the mean metallicity.
 If host galaxy stellar masses, star formation rates or dark matter
halo masses (and hence clustering) of GRB hosts evolve with redshift,
then the interpretation of burst rate as a proxy for star formation
rate history will need to be revisited and refined.

GRB\,090423 provides a case study at high redshifts,
potentially probing significant evolution in environment from $z<1$
burst hosts. Given that the GRB appears to be the result of a second
generation or later starburst at such early times, does it occupy a
more massive galaxy than typical of lower redshift bursts, a vigourous
starburst, a mature system that has exhausted its gas supply or a
young one with a large reservoir of processed molecular gas available
for later star formation?

One test of the latter hypothesis has become possible due to the broad
bandwidth and high sensitivity of the current generation of millimetre
telescopes. At $z=8$, the rest-frame infrared CO(3-2) emission line (a
good probe of molecular gas mass at low redshift) is shifted into the
observed 7\,mm band. The 4\,GHz bandwidth available using the new CABB
correlator at the Australia Telescope Compact Array (ATCA) enables a
volume with line-of-sight extent $\Delta z=0.9$ to be observed in a
single observation, easily encompassing the redshift uncertainty from
optical/near-infrared observations. Given the high redshift of this
and similar sources, carbon monoxide and other far-infrared emission
lines (notably [CII] at 1898\,GHz) may well prove the most accessible
diagnostics of the precise redshift of the GRB host and any galaxies
sharing a large scale structure surrounding it.

In this paper, we describe an investigation to study the CO(3-2)
emission line properties of the GRB host galaxy and neighbouring
galaxies in its immediate environs. We structure the paper as follows:
In section \ref{sec:obs} we describe our observations, targeted on the
host galaxy of GRB\,090423 and searching for CO(3-2) emission. In
section \ref{sec:grbhost} we present limits obtained on the properties
of the gamma ray burst host. In sections \ref{sec:contsources} and
\ref{sec:lineemitter} we present the properties of additional sources
detected in our data, first in continuum and then in molecular line
emission. Finally in sections \ref{sec:discussion} we discuss the
interpretation and implications of our results, before presenting our
conclusions in section \ref{sec:conc}.

All optical and near-infrared magnitudes in this paper are quoted in
the AB system \citep{1983ApJ...266..713O}.  We adopt a $\Lambda$CDM
cosmology with ($\Omega_{\Lambda}$, $\Omega_{M}$, $h$)=(0.7, 0.3,
0.7).

\section{Observations}
\label{sec:obs}

Observations were carried out over the period 2010 Mar 22-26th, using
the 7mm-band receivers at the Australia Telescope Compact
Array\footnote{Observations associated with programme
C2153-2009OCTS.}. The 4\,GHz bandwidth CABB Correlator was used, with
a small overlap between the two IFs, allowing continuous frequency
coverage in the range 36.075-39.723\,GHz, at 1\,MHz resolution. A
single pointing was observed at
09$^\mathrm{h}$55$^\mathrm{m}$33$\fs$29
+18$\degr$08$\arcmin$50$\farcs$80, placing the GRB Host Galaxy ten
arcseconds north of the phase centre. The primary beam FWHM of the
ATCA at 38\,GHz is 75$\arcsec$.

GRB\,090423 is well north of the optimal declination for the ATCA. As
a result, observations were split into 5 tracks and the source was
observed at elevations $>$30$\degr$ for 4.9hrs per track. To ensure
good uv-coverage and sensitivity at millimetre wavelengths,
observations were taken in the compact H168 configuration, yielding a
typical synthesised beam of $7.8\arcsec\times 6.3\arcsec$ at
half-power beam-width.

Bandpass calibration was performed on PKS\,0537-441 at the start of
every track, and nearby phase calibrator PKS\,0953+254 was observed at
10-15 minute intervals throughout the observations. The same source was
used to check the array pointing solution once an hour. PKS\,1934-638 (the
standard flux calibrator at the ATCA) was observed at the end of every
track to provide primary flux calibration.

Observations were scheduled at night in order to obtain the most
stable possible conditions for 7mm band observations. Conditions were
generally good throughout the observations, but data taken at
elevations $<30\deg$ were discarded due to a combination of increased
atmospheric opacity and less stable conditions at the start and end of
each night. Images were generated using natural weighting. The
measured RMS noise at the phase centre in the final images was
typically 74\,$\mu$Jy beam$^{-1}$ at 37\,GHz and 78\,$\mu$Jy
beam$^{-1}$ at 39\,GHz in 2\,MHz (16\,km\,s$^{-1}$) channels.

A multi-frequency (4GHz bandwidth) synthesis image centred at
37.887\,GHz was also generated from the data, reaching a sensitivity
limit (RMS) of 3.1\,$\mu$Jy beam$^{-1}$.

\section{Constraints on the $\mathbf{z\sim8}$ GRB Host}
\label{sec:grbhost}

The position of the $z\sim8$ GRB\,090423 is known to sub-arcsecond
accuracy (from near-infrared imaging of the burst afterglow). The
stellar light from galaxies at these distances is believed to have a
typical half-light radius of $\sim$0.3\,kpc (0.06$\arcsec$), based on
photometrically selected samples \citep{2010ApJ...709L..21O}. The beam
size of our ATCA observations is two orders of magnitude greater than
this, and any emission from the the GRB host galaxy itself is likely
to be contained within a synthesised beam centred at the burst
location (and probing a region $40\times30$\,kpc in extent). Our
initial analysis focused on the properties of our data at this
location.

No continuum source was seen at the GRB location in the
multi-frequency synthesis image, allowing us to place an upper limit
on the 37.9\,GHz continuum flux of this source of
$S_\mathrm{8mm}$=9.3\,$\mu$Jy (3\,$\sigma$, assuming the source is
unresolved). This observed frequency corresponds to the rest frame
850\,$\mu$m band and so probes the warm dust in the system. We use the
prescription of \citet{2008A&A...491..173A} to calculate an upper
limit on the dust mass at $z=8.2$, assuming the dust emission can be
modeled by a modified black body and accounting for the heating effect
of the cosmic microwave background at this redshift.  A number of
significant assumptions are required to make this calculation. Most
crucially, the dust coefficient and emissivity index are fixed at
local values, which may not be appropriate for high redshift sources,
particularly young and unevolved sources which are likely to be
dominated by supernova-processed dust grains
\citep[see][]{2004Natur.431..533M,2007ApJ...661L...9S}.  There is also
a weak dependence on the assumed size of the source. In the absence of
further data, we assume a source extent of 1\,kpc, while noting that
increasing this by an order of magnitude has negligible effect on the
derived mass.

 \begin{figure}
\includegraphics[width=0.99\columnwidth]{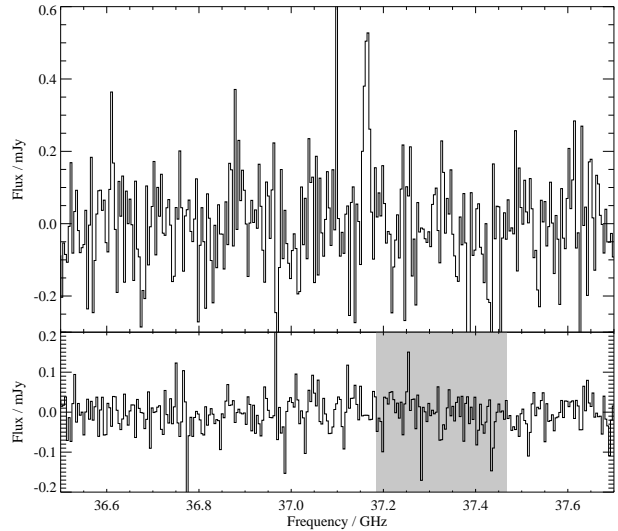}
\caption{The spectrum of our emission line source (above, see section
\ref{sec:lineemitter}), compared with the non-detection spectrum
extracted at the location of the GRB host galaxy (below, section \ref{sec:grbhost}) and assuming
the latter is a point source. Data are binned to 4\,MHz
(32\,km\,s$^{-1}$) resolution. The shaded region indicates the
expected frequency range for CO(3-2) emission from the GRB\,090423
host galaxy, assuming a 1\,$\sigma$ error range on the
optical/near-infrared redshift. The spectra shown span
10,000\,km\,s$^{-1}$ in velocity space and represent only 35\% of our total
observed frequency range. The effect of primary beam attenuation on
channel noise at the line emitter location is clearly visible.
  \label{fig:spectra}}
\end{figure}

Our derived upper limit on the dust mass in this source is
$8.0\times10^{8}$\,M$_\odot$ at 40K. Given the typical stellar masses
of GRB host galaxies at low redshift (see section
\ref{sec:discussion}) our non-detection of the source in dust
continuum is unsurprising.  The mean 850\,$\mu$m (observed) flux measured
by \citet{2004MNRAS.352.1073T} for a sample of 21 GRB hosts at
$0.45<z<3.42$ implies an average dust mass of only $\sim5\times10^{7}$\,M$_\odot$
for the lower redshift GRB host population
using the same prescription applied here. For thermal dust emission across a
reasonable range of temperatures, and for a typical low redshift GRB
host galaxy, our constraint implies only that the dust mass does not
exceed the stellar mass.
 
In addition to probing the dust continuum of the source, our
observations are sensitive to CO(3-2) line emission spanning the
redshift range $7.7<z<8.6$ -- easily encompassing the $\Delta z=0.13$
uncertainty in the optical redshift estimate for the GRB host. We have
examined our datacube carefully for such emission. Low luminosity
galaxies at high redshift have been observed to have lower line widths
than most submillimeter galaxies. Indeed. the only two known $z\sim5$,
non-AGN, CO-emitting galaxies at have line widths of
$\sim160$\,km\,s$^{-1}$ \citep{2010arXiv1004.4001C} and
$\sim110$\,km\,s$^{-1}$ \citep{2008ApJ...687L...1S} which may suggest
that the molecular gas in these sources  does
not form a rotationally-supported disk. With these narrow lines in
mind as a model, we select 150\,km\,s$^{-1}$ as a plausible velocity
width for any emission line in our distant source.

In order to optimise our sensitivity to emission with this and a
variety of both broader and narrower velocity widths, the spectral
datacube was rebinned at resolutions of 2, 4, 8 and 16\,MHz and each
rebinned cube was analysed independently. Despite this, a spectrum
taken through the data cube at the location of the GRB host galaxy
shows no evidence for line emission at the sensitivity limit of the
data at any spectral resolution (see figure \ref{fig:spectra}). This
places an upper limit on the velocity integrated line width of the
CO(3-2) transition (expected at 37.45$^{+0.29}_{-0.24}$\,GHz) of
$S_{CO}\Delta v<$0.026\,Jy\,km\,s$^{-1}$ (3\,$\sigma$) for a
150\,km\,s$^{-1}$ wide emission line.

The estimation of molecular gas mass from CO line intensity requires
the application of an empirical calibration for
M$_\mathrm{gas}$/$L'_\mathrm{CO}$. For ease of comparison with other authors, 
we assume that the molecular gas in
this source is optically thick and, apply the standard conversion
factor M$_\mathrm{gas}$/$L'_\mathrm{CO}$=0.8\,M$_\odot$\,(K
km\,s$^{-1}$)$^{-1}$, derived for ULIRGS and SMGs at intermediate
redshifts and believed to be constant for such sources between $z=0$
and $z=4$ \citep[see][]{2005ARA&A..43..677S}. Using this conversion
we determine an upper limit on the
molecular gas at $z=8.23$ of M(H$_2$)$<4.3\times10^{9}$\,M$_\odot$
(3\,$\sigma$).

We note that the conversion has not yet been demonstrated for low mass
galaxies forming stars at a relatively low rate, at significantly
sub-solar metallicities metallicities, or at early times, for which a
higher conversion factor may be more appropriate. Applying the
Galactic conversion factor,
M$_\mathrm{gas}$/$L'_\mathrm{CO}$=4.6\,M$_\odot$\,(K
km\,s$^{-1}$)$^{-1}$, would yield limits a factor of 5.75 times weaker
than those presented here. The true value of the conversion factor is
unknown but likely to lie somewhere between these extremes \citep[see
discussion in][]{2008ApJ...680..246T}.

\section{Radio Sources in the GRB Field}

\subsection{Continuum Sources}
\label{sec:contsources}

We quantify the number density of continuum sources in our field,
considering a circular region extending from the pointing centre to
the radius at which primary beam attenuation limits sensitivity to
40\% of its maximum value (47$\arcsec$), yielding a survey area of
$5.4\times10^{-4}$\,deg$^{-2}$. We detect a single unresolved
5\,$\sigma$ source in our 4\,GHz-bandwidth continuum image, with a
peak flux $S_\mathrm{8mm}=15.9\pm3.2$\,$\mu$Jy, determined by fitting
an Gaussian point source to the image data. The source lies 33
arcseconds to the south-east of the GRB host galaxy. Two other
features in the continuum map occur at 3\,$\sigma$ relative to the
local noise level.

In each case, several faint galaxies are visible at the source
location in available deep $J$ and $H$ band imaging of the field (see
section~\ref{sec:lineemitter}) but the large beam size of the ATCA at
39\,GHz prevents secure identification of the correct near-infrared
counterpart (if any) or determination of an accurate redshift.  None
of the three continuum sources show evidence for strong line emission
in the frequency range observed.

\subsection{A Molecular Line Emitter}
\label{sec:lineemitter}

In principle we are sensitive to any emission line source appearing in
the telescope primary beam and in our 36-40\,GHz frequency range. In
practice, this region of the spectrum is largely devoid of significant
lines from Galactic or low redshift sources
\citep{1979ApJS...41..451L}. The only strong features seen from
Galactic sources in this frequency range (i.e. emission lines
without comparable-strength neighbouring lines) are high order
transitions in methanol masers, which are usually highly
polarised and associated with high mass star formation regions.

There are only three credible extragalactic interpretations of an
emission line occurring at these high frequencies. We are sensitive to
CO(1-0) line emission occurring at $z=2.05\pm0.15$, CO(2-1) at
$z=5.10\pm0.30$ and CO(3-2) line emission at $z=8.15\pm0.45$. 

The data cube obtained contains spectral information for any source
lying within the primary beam of the ATCA. We rigorously inspected our
data cube for emission line candidates lying within 47$\arcsec$ of the
pointing (and hence phase) centre. 
As in the case of the GRB host galaxy, we consider the data rebinned
in frequency on scales ranging from 16 to 126\,km\,s$^{-1}$ per
spectral resolution element, searching both for particularly narrow
line emission and for weaker but broader emission lines. Clear
detections at the same spatial location in three or more adjacent
channels were required to identify an emission line candidate. While pattern noise in the
reduced images often creates strong peaks in the noise distribution,
these are limited to one spectral channel, and noise peaks in adjacent
channels appear spatially offset from one another.

We identify a single strong emission line source in our spectral data
cube, centered at 37.164\,GHz as shown in figure \ref{fig:spectra}. The
best fitting Gaussian line profile has a peak flux of $0.58\pm
0.12$\,mJy and a velocity full width at half-maximum of
96\,km\,s$^{-1}$, yielding an integrated line flux $S_{CO}\Delta
v=0.059\pm0.012$\,Jy\,km\,s$^{-1}$, accounting for primary beam
attenuation\footnote{which reduces the measured flux at this position
to 42.5\% of its true value}. The line flux is not significantly
polarised (although we note that at this low signal to noise there is
large flux uncertainty in each polarisation). The source, located at
09$^\mathrm{h}$55$^\mathrm{m}$35$\fs$96
+18$\degr$08$\arcmin$24$\farcs$7, is not detected in 8mm continuum
emission.

  \begin{figure*}
\includegraphics[width=0.85\columnwidth]{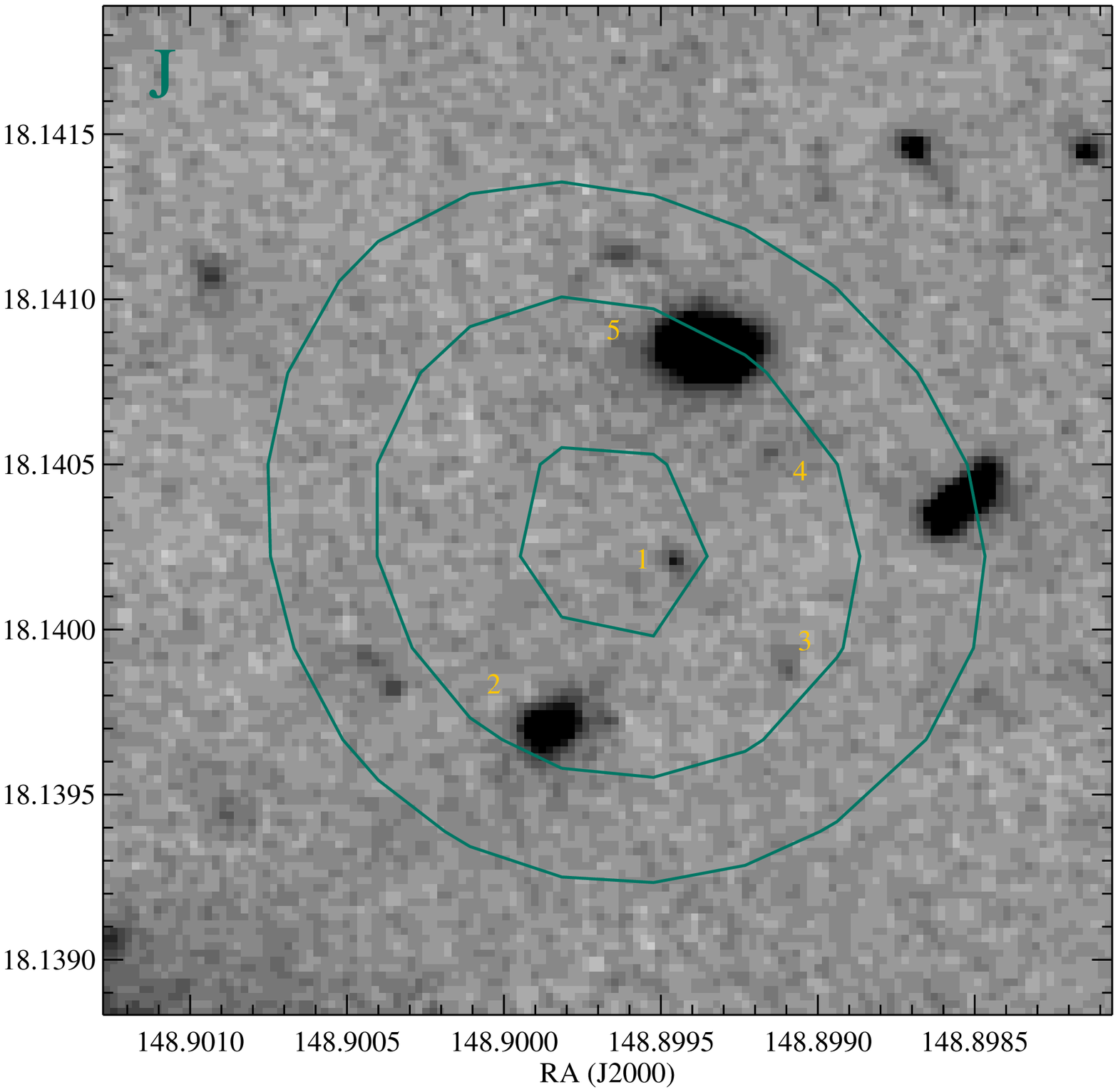}
\hspace{0.5cm}
\includegraphics[width=0.85\columnwidth]{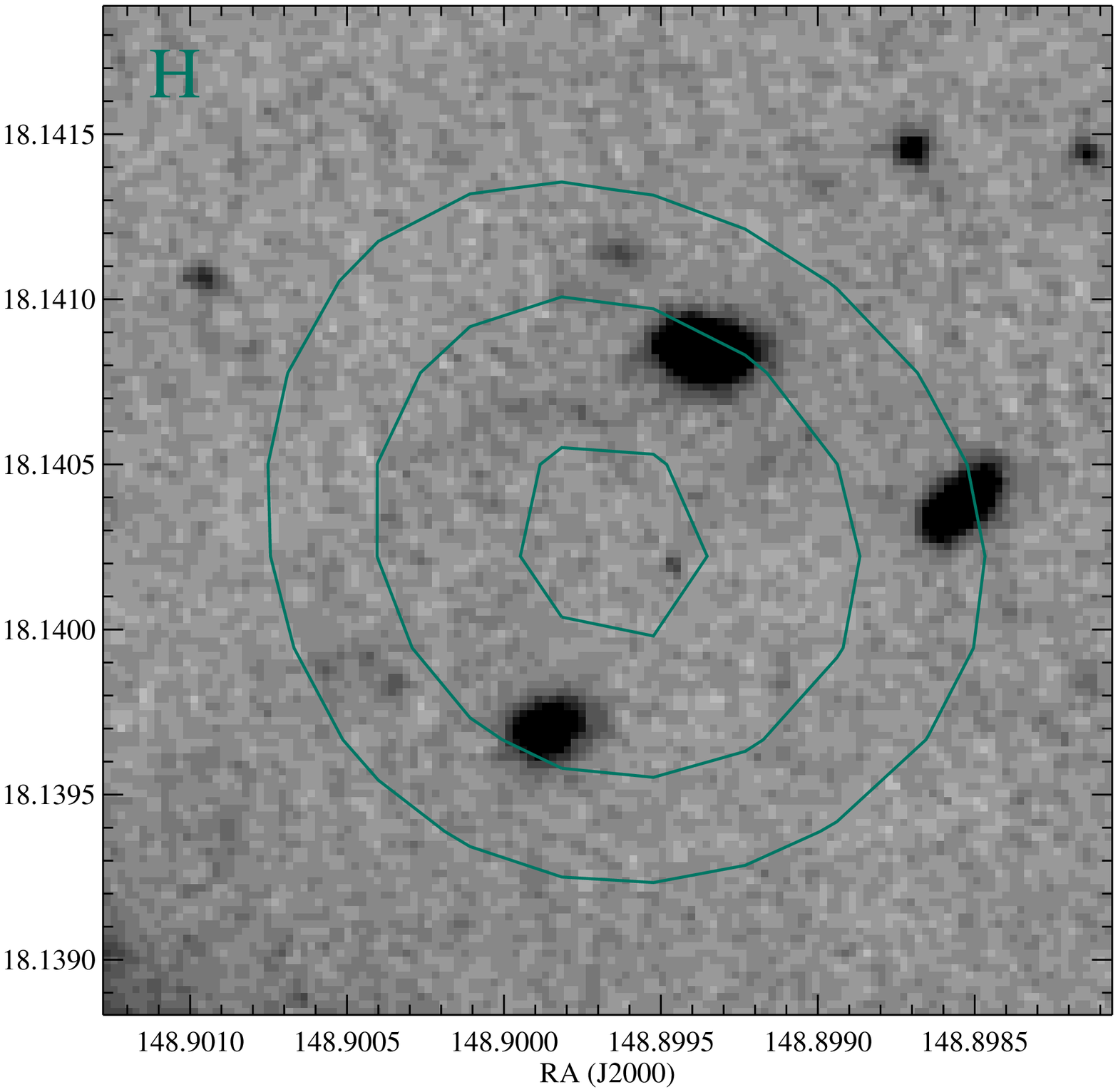}
\caption{Deep near-infrared imaging in the F125W (J) band and F160W
(H) band, overlayed with contours showing the spatial position of CO
line emission, integrated across the 22\,MHz width of the line
emission. Contours mark 5, 7.5 and 10\,$\sigma$ levels relative to the
noise in the line-integrated image. Images are 12 arcseconds on a side
and oriented with north uppermost. As discussed in section
\ref{sec:lineemitter}, the synthesised beam of our observations at
this frequency is $8\farcs00\times6\farcs45$ and the pointing is
accurate to approximately half a beam (roughly corresponding to the
second contour). It is apparent that a number of plausible
counterparts exist for the line emitter with the five best-detected
labelled in the J band image
  \label{fig:images}}
\end{figure*}

As discussed above, three plausible redshifts exist for this source:
$z=2.10$, 5.20 and 8.31, corresponding to the first three rotational
transitions of the carbon monoxide molecule. Due to the offset between
redshift, luminosity distance and intrinsic line flux for these three
transitions, our observations reach comparable mass sensitivities in
all three lines: 1.0, 1.1 and 0.98 $\times10^{10}$\,M$_\odot$ in
molecular gas respectively (assuming
M$_\mathrm{gas}$/$L'_\mathrm{CO}$=0.8 as discussed in
section~\ref{sec:grbhost}).

Since the galaxy almost certainly does not fill our $7.8\arcsec\times
6.3\arcsec$ beam, a detailed calculation of the dynamical mass is not
possible with the current data, and, in fact, it is not clear whether
the line width is actually diagnostic of ordered rotation in sources
of this type. However, while narrow, the emission line velocity width
is consistent with the calculated gas mass if the molecular gas is
distributed in a virialised disk one arcsecond in diameter at $z=8$
(0.6\,arcseconds at $z=2$), seen close to face-on.

Given the hierarchical model for galaxy formation, and in the absence
of additional information on large scale structure in this field, our
\textit{a priori} most likely interpretation has to be that this is an
intermediate-mass star forming galaxy at $z=2.10$. Such a galaxy would
typically be easily detected in deep optical \citep[$I\approx24$ in
similar mass examples, see][]{2009MNRAS.399..121C} and near-infrared
data, although the presence of significant amounts of dust could lead
to the absence of a counterpart (as is the case for submillimetre
luminous galaxies). Higher redshift sources would typically be both
fainter and bluer in the near-infrared.

In figure \ref{fig:images} we show existing deep near-infrared imaging
of the region from {\em HST}/WFC3 (P.I. Tanvir, analysis in
prep). This data reaches a limit of $J\sim28.5$ (AB, $3\,\sigma$) in a
$1\arcsec$ aperture and was obtained in order to study the host
galaxy of GRB\,090423. Near-infrared data of this depth and quality is
rare. Figure \ref{fig:images} demonstrates the high-surface density of
faint sources in such data, and hence the difficulty of assigning an
unambiguous counterpart to the line emitter. The ambiguity is
further heightened by uncertainties in the pointing accuracy of the
ATCA. At millimetre wavelengths, the telescope is used in reference
pointing mode, with a nearby calibrator (PKS\,0953+254 in this case)
used to check and refine the pointing solution once an hour throughout
observations. Typical corrections at the end of each hour's
observations were of order a few arcseconds, and rarely exceeded half
the size of the synthesized beam. 

Given this uncertainty, there are five near-infrared sources,
well-detected in both bands, which may represent a counterpart for the
line emission source, as well as a number of low-surface brightness
features at the noise level of the data. The $J-H$ colours of these
sources are insufficient to provide a redshift estimate, although we
note with interest that the faint unresolved point source closest to
the centre of the line emission (marked `1' in figure
\ref{fig:images}) shows very blue near-infrared colours ($J=27.4$,
$J-H=-0.83\pm0.15$), a trait typical of high redshift ($z>5$) galaxies
\citep[e.g.][]{2005MNRAS.359.1184S,2010ApJ...708L..69B} and difficult
to obtain for normal stellar populations at lower redshift.

Further data will be needed to unambiguously confirm the redshift of
this source, and in the absence of such data, a tentative identification
as CO(1-0) emission from a faint $z=2.10$ starburst galaxy remains the
most likely interpretation.

\section{Discussion}
\label{sec:discussion}

\subsection{Constraints on the GRB Host Galaxy}
\label{sec:dis_host}

A typical L$^\ast$ Lyman break galaxy at $z\approx7-8$ is believed to
have a stellar mass of a few $\times 10^9$\,M$_\odot$
\citep{2010ApJ...713..115G,2010ApJ...708L..26L}, although considerable
uncertainties (primarily stellar population synthesis model used,
metallicity, photometric uncertainty and the deblending of faint,
confused sources) remain in this calculation. This is very comparable
to the median stellar mass determined for GRB host galaxies at $z<1.2$
\citep[$1.3\times10^9$\,M$_\odot$, ][]{2010MNRAS.tmp..479S}, and
little evolution in this is seen to much higher redshifts either directly
\citep{2009ApJ...691..182S} or in semi-analytic simulations tuned to match
the observed distribution at $z=1-6$ \citep{2010arXiv1005.4036C}.

Our constraint on the mass of molecular gas in the host galaxy of
GRB\,090423 requires M$_\mathrm{H2}$ to be within
a factor of a few of this expected stellar mass for GRB hosts,
suggesting either that the GRB host is less massive than those typical
at lower redshift or that its star formation is already well advanced,
i.e. that it must be at least a third of the way through its starburst
process, assuming 100\% efficiency in the conversion of gas to stars.

This tight limit is sensitive to gas not only in the host galaxy
itself, but in the $40\times30$\,kpc region encompassed by the
synthesised beam of the observations. As a result the upper limit
includes gas infalling on or recently ejected from a star forming
galaxy, and in any close neighbours. As such it is somewhat surprising
that a galaxy that is likely the progenitor of a much more massive low
redshift system has such a limited fuel supply available for later and
ongoing star formation. While no data is currently available on the
current star formation rate of this source, it is instructive to
consider a crude model in which the host galaxy is typical of lower
redshift GRB hosts (i.e. M$\ast\approx 10^9$\,M$_\odot$) and star
formation began at $z\approx12$ (during the epoch of reionisation),
yielding a a stellar population age of 240\,Myr and a continuous 
star formation rate of 4\,M$_\odot$\,yr$^{-1}$
 \citep[comparable to the median of 3.8\,M$_\odot$\,yr$^{-1}$ found in the
GRB host sample of][]{2010MNRAS.tmp..479S}.  In this model, the host
galaxy will exhaust its fuel supply by $z\approx5.7$ at the current
rate of star formation. If the host is less massive or star formation
began earlier (and hence the average star formation rate is lower) the
star formation epoch may extend to lower redshifts.

We note that this conclusion is drawn from the assumption of an
optically-thick thermalised, metal-enriched gas, and from a CO to
H$_2$+He gas mass conversion factor that is calibrated on the most
massive starbursts at intermediate redshifts. These assumptions may
well be flawed on a number of levels.  As mentioned in
section~\ref{sec:grbhost}, the applicability of a simple conversion
(widely known as the X-factor) from CO(1-0) luminosity to a gas mass
has not been demonstrated at higher redshifts, or in galaxies with
sub-ULIRG star formation rates. The low
M$_\mathrm{gas}$/$L'_\mathrm{CO}$ ratio in ULIRGs is believed to arise
from the presence of a galaxy-wide starburst in which the bulk of the
interstellar medium is both dense and irradiated by star formation
\citep{1997ApJ...478..144S}. If the GRB host galaxy at $z\sim8$ is,
like those at intermediate times, similar in character to $z\sim5$
Lyman break galaxies in terms of mass, metallicity and star formation
rate, then it is interesting to note that such systems do indeed seem
to be galaxy-wide starbursts with high specific star formation rates
\citep{2007MNRAS.377.1024V} and no evidence that the UV-luminous knots
observed are merely super-starclusters embedded in a more massive,
cooler medium \citep{2010arXiv1007.3989D,2010arXiv1007.0440S}. While
this may argue in favour of the ULIRG-like conversion ratio used in
this paper, the low metallicity and comparitively low star formation
rates of such sources may have the effect of increasing the
ratio. This significant source of uncertainty must necessarily be
addressed by later studies.

Although GRB\,090423 is believed to
be a population II event and thus indicative of a metal enriched star
forming population, the fraction of relatively pristine H$_2$ gas in
the outer regions of a galaxy at early times is likely to be higher,
particularly well outside the star forming disk. Our observations
cannot rule out the presence of a large reservoir of such gas, either
in the host galaxy or in filaments surrounding it since this would
likely not be reflected in the measured CO luminosity. 

Similarly, the conversion assumes a known ratio of emission in the
higher order rotational lines to that in the CO(1-0) fundamental
transition. At high redshifts, the cosmic microwave background
provides a minimum temperature for a typical galaxy, pushing the
rotational excitation ladder of carbon monoxide to higher levels and
hence reducing the measured flux in the lower excitation states. The
effect of this is seen in the recent models of
\citet{2009ApJ...702.1321O}, which indicate that the highest surface
density of sources to a given flux limit is seen in CO(3-2) emission
at low redshift and in CO(5-4) emission by $z=5$. Future follow-up of
this and similar sources with ALMA (which, while having a more limited
field of view, will access higher order CO transitions) may yield an
indication of this effect and prove more successful in determining the
precise source redshift and gas mass.

\subsection{Large Scale Structure at High Redshifts}
\label{sec:dis_lss}

As mentioned in section~\ref{sec:lineemitter}, our deep datacube
probes cosmological volume in each of three redshift
ranges. Considering the region in which our flux sensitivity is
$>$40\% of that at the pointing centre yields a field of view of
$5.4\times10^{-4}$\,deg$^{-2}$, while the 4\,GHz bandwidth yields a
line of sight depth of $\Delta z=0.29$, 0.59 and 0.88 at $z=2$, 5 and
8 respectively.

  \begin{figure}
\includegraphics[width=0.98\columnwidth]{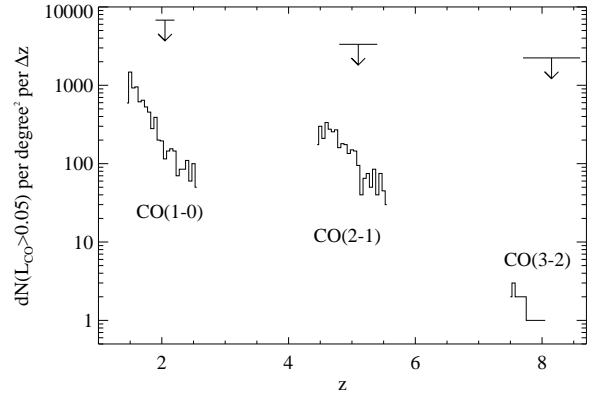}
\caption{Our upper limit on the number density of line emission
sources satisfying $L'_\mathrm{CO}>0.05$\,Jy\,km\,s$^{-1}$, given the
velocity integrated flux of our single detected line emitter
(i.e. assuming that we detect 1 or fewer sources in each redshift
range). Also shown are predicted number densities of sources
satisfying the same constraint in each emission line from the
semi-analytic models of \citet{2009ApJ...703.1890O} as discussed in
section \ref{sec:dis_lss}. Since these models are based on a section
through the millennium simulation, they are subject to small number
statistics in the predicted number densities, particularly at the
highest redshifts.
\label{fig:models}}
\end{figure}

Given a single emission line source (with redshift yet to be
determined) we are able to place an upper limit on the surface density
of CO-luminous galaxies in each of these redshift ranges. Our
constraints are shown in figure \ref{fig:models}, together with the
predicted surface densities of sources with
$L'_\mathrm{CO}>0.05$\,Jy\,km\,s$^{-1}$ (i.e. to a flux limit
matching our detection of a single possible line) determined from
the \textit{S-cubed} simulations of \citet{2009ApJ...703.1890O}. These
analyse a conical section through the Millennium Simulation, applying a
semi-analytic prescription to calculate the predicted CO rotation
emission line excitation ladder and H$_2$ gas properties associated
with about $2.8\times10^8$ dark matter
haloes\footnote{http://s-cubed.physics.ox.ac.uk/s3\_sax/sky}. As figure
\ref{fig:models} demonstrates, very few $z=8$ galaxies in the
simulation would satisfy our flux selection limit in the CO(3-2) line
(the simulation field of view at $z=8$ is about
$4\times4$\,degrees$^2$), and hence the uncertainty on the predicted
number densities is dominated by small number statistics.

Interestingly, regardless of which line identification is considered
most likely, our detection of a CO-emitting source is something of a
surprise.  Given these simulations, and in the absence of lensing, we
might expect to have detected either a $z=2.1$ CO(1-0) emission line
or a $z=5.2$ CO(2-1) emission line of this strength in one of 35 blind ATCA
pointings. It is possible that one of the intermediate redshift galaxies visible
in figure \ref{fig:images} is a lensing source, boosting the flux of a
$z=2$ or $z=5$ background galaxy, or that we have chanced across rare
large scale structure at these redshifts.

The probability of a randomly selected pointing detecting a $z=8$
galaxy in CO emission is considerably lower (approximately one in two
thousand). However this field is not randomly selected. In the
unlikely event that follow-up investigations confirm the very blue
($J-H=-0.8$) faint galaxy as a $z=8$ near-infrared counterpart to the
emission line source, this field would host both a Gamma Ray Burst and
a more massive gas-rich source (separated by a projected distance of
200\,kpc), representing a highly significant overdensity above the
typical galaxy distribution at this redshift and suggesting that GRBs
are indeed effective tracers of large scale structure at high
redshift. While constraints from a single GRB and associated field remain weak,
such a detection might argue against the suppression of star formation
in low mass haloes seen in warm dark matter models, as discussed by
\citet{2005ApJ...623....1M} which are expected to yield both
fewer high redshift GRBs and fewer star forming galaxies associated
with them.

Even by $z=8$, the effects of cosmic variance can be
significant. \citet{2010arXiv1005.3139I} have suggested a two order of
magnitude variation in the density of 0.5\,Mpc$^3$ regions at this
redshift, based on large N-body simulations probing structure
formation during reionisation. Given the significant expenditure of
telescope time necessary to obtain deep imaging suitable for exploring
the $z\sim8$ Universe, exploring underdense regions is likely to be
both disappointing and uninformative. Following up GRB fields (should
additional $z>6$ bursts be identified) may well be the most efficient
method for identifying the high density regions that are the sites for
later massive galaxy and cluster formation.

\section{Conclusions}
\label{sec:conc}

We have observed the host galaxy of GRB\,090423 and its immediate
environs at radio frequencies (36.1-39.7\,GHz), in order to search for
molecular gas emission in the CO(3-2) line. Our main conclusions can
be summarised as follows:

\begin{enumerate}

\item No continuum source was seen at the GRB location in the
multi-frequency synthesis image, allowing us to place an upper limit
on the 37.9\,GHz flux of this source of $S_\mathrm{8mm}$=9.3\,$\mu$Jy
(3\,$\sigma$, assuming the source is unresolved).

\item No line emission is detected from the GRB\,090423 host
galaxy. Applying standard conversion factors, we determine an upper
limit on the molecular gas in this $z=8.23$ source of
M(H$_2$)$<4.3\times10^{9}$\,M$_\odot$ (3\,$\sigma$).

\item This tight limit on the fuel for star formation suggests either
that the GRB host is less massive than those seen at lower redshift, or
that its starburst is well advanced. We caution that this conclusion 
depends critically on the assumptions required to calculate gas masses,
which are not yet well calibrated at high redshift.

\item We identify a single strong emission line source in our data,
centered at 37.164\,GHz and 44 arcseconds to the south-west of the
GRB. The emission has a peak flux of $0.58\pm 0.12$\,mJy and a
velocity full width at half-maximum of 96\,km\,s$^{-1}$. This
corresponds to a molecular gas mass of 1.0, 1.1 or 0.98
$\times10^{10}$\,M$_\odot$ if interpreting the line as CO(1-0) at
$z=2.1$, CO(2-1) at $z=5.2$ or CO(3-2) at $z=8.3$
respectively. Efforts to identify and characterise a counterpart to
this emission line source are continuing.

\end{enumerate}

\section*{Note added in Press}
  Flux limits given in sections 2 and 3 are obtained from data that
  has been `cleaned' using {\sc MIRIAD}. Further inspection and consultation
  with ATCA staff suggests a less vigorous cleaning strategy may be
  appropriate. This would result in an effective increase in the image
  noise, and hence quoted flux and mass constraints, by a factor of
  two ($\sim$3 if cleaning is omitted completely from the data
  reduction). Quoted fluxes for the line emission candidate are
  unaffected, and there is no substantive impact on our discussion or
  science conclusions.

\section*{Acknowledgments}
ERS gratefully acknowledges support from the UK Science and Technology
Facilities council.  Based on data obtained at the ATCA as part of
program C2153. The Australia Telescope Compact Array is part of the
Australia Telescope which is funded by the Commonwealth of Australia
for operation as a National Facility managed by CSIRO. The WFC3 data
referenced in section \ref{sec:lineemitter} are associated with 
HST Proposal 11189 and we thank the co-investigators of that project for
kindly allowing us access to the data here. The authors
would like to thank Mark Birkinshaw and Matt Lehnert for useful
discussions during the preparation of this manuscript. We also thank
Ben Humphries and Shari Breen for their assistance as Duty Astronomers
during our observations.

\bsp

\label{lastpage}

\end{document}